\documentclass[utf8]{frontiersSCNS} %
\usepackage{url,hyperref,lineno,microtype,verbatim}
\usepackage{todonotes}
\usepackage[onehalfspacing]{setspace}
\usepackage{multirow}

\usepackage{ulem} %

\newtheoremstyle{query}%
{}{}
{\color{red}}
{}
{\sffamily\bfseries}{:}{12pt}
{}
\theoremstyle{query}
\newtheorem{aq}{Author Query/Comment}

\newcommand{\AQ}[1]{{\textcolor{black}{#1}}}

\newcommand{\baq}{\begin{aq}}
\newcommand{\eaq}{\end{aq}}

\def\keyFont{\fontsize{8}{11}\helveticabold }
\def\firstAuthorLast{Panchea {et~al.}} 
 \def\Authors{Adina M. Panchea\,$^{1,\star,a}$, Dominic Létourneau\,$^{1,a}$, Simon Brière\,$^{2,a}$, Mathieu Hamel\,$^{2,b}$, Marc-Antoine Maheux\,$^{1,a}$, Cédric Godin\,$^{1,a}$, Michel Tousignant\,$^{2,a}$, Mathieu Labbé\,$^{1,c}$, François Ferland\,$^{1,a}$, François Grondin\,$^{1,b}$, François Michaud\,$^{1,a}$}
 \def\Address{$^{1}$IntRoLab - Intelligent, Interactive, Integrated and Interdisciplinary Robotics Lab, Interdisciplinary Institute for Technological Innovation (3IT), Department of Electrical Engineering and Computer Engineering,\\ 
$^{2}$Research Center on Aging, School of Rehabilitation,\\
Université de Sherbrooke, Sherbrooke, Quebec, Canada\\ 
$^{a}$ $\{$firstname.lastname$\}$@usherbrooke.ca \\
$^{b}$ $\{$firstname.lastname2$\}$@usherbrooke.ca \\
$^{c}$ $\{$mathieu.m.labbe$\}$@usherbrooke.ca 
}
\def\corrAuthor{Corresponding Author}
\def\corrEmail{adina.panchea@usherbrooke.ca}
\begin{document}
\onecolumn
\firstpage{1}
\title[OpenTera and Solutions to COVID-19 Challenges in Care Facilities]{OpenTera: A Microservice Architecture Solution for Rapid Prototyping of Robotic Solutions to COVID-19 Challenges in Care Facilities
} 
\author[\firstAuthorLast ]{\Authors} 
\address{} 
\correspondance{} 
\extraAuth{}
\maketitle
\begin{abstract}
As telecommunications technology progresses, telehealth frameworks are becoming more widely adopted in the context of long-term care (LTC) for older adults, both in care facilities and in homes. 
Today, robots could assist healthcare workers when they provide care to elderly patients, who constitute a particularly vulnerable population during the COVID-19 pandemic. 
Previous work on user-centered design of assistive technologies in LTC facilities for seniors has identified positive impacts. 
The need to deal with the effects of the COVID-19 pandemic emphasizes the benefits of this approach, but also highlights some new challenges for which robots could be interesting solutions to be deployed in LTC facilities.
\AQ{This requires customization of telecommunication and audio/video/data processing to address specific clinical requirements and needs.}
This paper presents OpenTera, an open source telehealth framework, \AQ{aiming to facilitate} prototyping of such solutions \AQ{by software and robotic designers}. 
Designed as a microservice-oriented platform, OpenTera is an end-to-end solution that employs a series of independent modules for tasks such as data and session management, telehealth, daily assistive tasks/actions, together with smart devices and environments, all connected through the framework. 
After explaining the framework, we illustrate how OpenTera can be used to implement robotic solutions for different applications identified in LTC facilities and homes, and we describe how we plan to validate them through field trials.
\tiny
\keyFont{ \section{Keywords:} telehealth, telecommunications framework, microservice architecture, long-term care facilities, COVID-19 challenges, mobile robot, teleoperation} 
\end{abstract}

\section{Introduction}

Telehealth is defined as the use of information and communication technologies (ICT) to extend healthcare service delivery over long distances \citep{chi2011}. 
Its success relies on the ability to maximize the quality of care provided remotely, i.e., to provide diagnosis and treatment of  equivalent quality  to what can be achieved by face-to-face or conventional means, and to maximize  benefits for  patients,  clinicians, and the healthcare system. 
It is thus able to improve patient outcomes and satisfaction, reduce the healthcare resources required for improved care, provide cost savings and more efficient use of healthcare resources, and reduce hospitalizations \citep{elder2013}. 
There are, however, significant barriers to the adoption and implementation of telehealth, such as cost, technological incompatibility, privacy and security issues, usability, acceptability, and education of healthcare personnel in the use of the associated technology \citep{elder2013,reginatto2012,kubitschke-cullen2010,goodwin2010,alwan-nobel2008,broens-veldw-2007}.
These barriers are intertwined, making adaptability of the technology a key element for success.  

As of September 2020, more than 32 million people have been confirmed to have been infected with COVID-19.\footnote{\url{https://coronavirus.jhu.edu/map.html}}  
\AQ{The most common symptoms of COVID-19 are fever, cough, and tiredness.} 
Moreover, in the absence of a vaccine or antiviral treatment, strict measures must be followed to prevent, contain, and attenuate the spread of the virus among  older adults, considered to be a high-risk population. 
These strict measures consist of good hygiene, social distancing, restrictions on social activity and travel, isolation and quarantine measures, and hospitalization among the exposed, suspicious, and positively tested population. 
In the context of long-term care (LTC) facilities for seniors, these measures are very challenging, and some of them are almost impossible to implement, since care workers play a critical role and have to work in challenging conditions to provide daily assistance to elderly people.%

As a result, the COVID-19 pandemic has drastically changed the landscape of telehealth barriers: people are now willing to try telehealth technologies, disregarding previous concerns. 
Videoconferencing solutions such as Zoom and Microsoft Teams have seen large increases in their users, and their integration into our work strategies is now becoming a reality.
\AQ{\cite{tavakoli2020} also describe how robotic, autonomous, and smart devices can be used to help front line workers in the healthcare system  during the COVID-19 crisis and discuss examples of how techniques from various areas of medicine, engineering, and science can help healthcare workers as well as society as a whole.}
However, some barriers remain with conventional videoconferencing solutions, which basically only establish an audio and video link between two or more participants. %
For instance, telehealth systems \AQ{may} also require 
easy installation and use, 
remote biometrics and remote control of devices (e.g., vital sign monitoring devices, pan--tilt--zoom (PTZ) cameras, and telepresence robots), streaming of multiple video feeds, and specialized graphical interfaces to improve situation awareness and reduce 
cognitive load.

In spite of the proliferation of videoconferencing solutions, adaptation of commercial videoconferencing platforms to address specific telehealth needs may be difficult or impossible to accomplish, depending on the priorities set by the providers of these solutions. 
To cope with these limitations, we chose to develop a customizable telecommunication platform to address the broad range of specifications and requirements of clinical and telehealth applications \citep{lepage-letourneau-hamel-2014,lepage-letourneau-hamel-2016}. 
To do so, we follow a holistic, bottom-up, iterative, and incremental approach starting from small and simple systems addressing specific needs and current practices, in situ, and designed to minimize changes to clinical practice, rather than  expecting  clinicians and patients to simply adapt to the technology.
Simplicity of use is a central requirement in these development efforts. 
More specifically, we did the following:
\begin{itemize}
\item In 2004, we started by developing a telerehabilitation platform integrating a commercially available videoconferencing system (Tandberg 550 MXP, using H.323 and H.263 codecs) providing two video feeds using the system's integrated PTZ camera, a touchscreen computer, and external biosensors, along with a homemade companion application providing an overview of all the available patients and easing the connection process for the clinician and the patient \citep{tousignant-moffet-2011,tousignant-boissy-2011,tousignant-marquis-2012,dechene-tousignant-2011}.
The data collected during a session included session information (which clinician with what patient for how long and what session type---physiotherapy, speech therapy, etc.), data from external sensors (oxygen saturation and heart rate), network performance (bandwidth, jitter, packet loss, and reported technical problems), and results from a satisfaction survey (for the clinician only), all recorded in a local database. 

\item In 2013, with the availability of inexpensive and general purpose embedded computers able to process audio, video, and data in real time, a new version of the platform, named Vigil2, integrated a software videoconferencing codec (H.264 codec for video, Speex for audio) to reduce cost and complexity and facilitate in-home deployment, making it an all-in-one solution not relying on external hardware codecs. That platform was also built to allow integration of new sensors (ECG) and network PTZ cameras.

\item In 2017, to address needs such as the ability to see the patient in different areas (kitchen, bathroom, bedroom, and staircases) and to further reduce the cost of the hardware platform by instead using  patient devices, we integrated mobile handheld audio--video devices (such as smartphones and tablets, using the integrated Internet browser supporting WebRTC technology), wearable sensors, and a telepresence robot with the revised platform, named TeraPlus.

\end{itemize}

Going through these development phases in close association with clinical trials, it became clear to us that a more generic Healthcare Internet-of-Things (H-IoT) platform, which could cover remote patient monitoring, videoconferencing (including for telerehabilitation, teleconsultation, and general telehealth applications), telepresence robotics, smart environments, and features required by the various projects and clinician needs, would provide key benefits for prototyping solutions adapted to telehealth needs.
Still, keeping up with software maintenance and upgrades of videoconferencing codecs, along with compatibility with a wide range of computing devices (computers, tablets, and smartphones, using different operating systems), requires a framework based on generic and widely adopted standards, rather than one using privately designed packages as in the previous iteration of the platform.
The lack of multiservice architecture solutions detached from a specific application \citep{napoli2019} or robotic platform \citep{cosar2020, Ferland2019} motivated our team to conduct research in this direction.
Therefore, we decided to develop OpenTera, an open source microservice telecommunication framework for telehealth applications based on what we have learned with previous experiments and platforms.
OpenTera is designed to provide useful, secure, and reliable telecommunication services independent of tier actors and online services.
Making it open source allows it to exploit standardized tools, avoids reinventing what is already available, and focuses on the addition of new capabilities and their integration.
This is similar to what the open source Robot Operating System (ROS) framework \citep{Quigley2009} accomplishes for the robotic community, and provides the opportunity to encourage other researchers to eventually contribute to the effort. 

The goal of this paper is to present OpenTera and its microservice architecture, and to describe how it can be beneficial in handling the COVID-19 pandemic, with an emphasis on its use in LTC facilities for older adults. 
\AQ{OpenTera is a framework intended for software engineers and roboticists to prototype and to deploy H-IoT applications without having to start from the beginning everytime.}
The organization of the paper is as follows. Section~\ref{sect2} presents OpenTera along with details of its microservice modules. 
Section~\ref{sect3} reports the needs and challenges identified with LTC facilities and homes during the COVID-19 pandemic, \AQ{outlining requirements that commercial videoconferencing systems do not address. Section}~\ref{sect4} describes \AQ{how OpenTera can be exploited in such contexts}, along with some considerations in deploying such solutions. 
Section~\ref{sect5} \AQ{concludes the paper}.

\section{O\lowercase{pen}T\lowercase{era}: A Microservice Telehealth/H-I\lowercase{o}T Platform}\label{sect2}

In modern computing, microservices are a software development technique in which an application is structured as a set of loosely coupled services. In contrast  to a monolithic system, which is formed as a single process by integrating all the interfaces and functionalities, the independent microservices offer the same functionalities and interfaces to the external user. However, the microservices are separated into multiple components that communicate with each other using standardized application programming interfaces (APIs) that are independent of the programming language. 
The most important microservice features are as follows \citep{dragoni2017}: 
\begin{enumerate}
\item[(a)] They can be independently and directly testable, in isolation and without affecting the whole system. 
\item[(b)] They can foster continuous integration to ease maintenance: for example, performing changes on a module does not require a reboot of the whole system. 
\item[(c)] They have their own containers with their own configuration of the deployment environment. 
\item[(d)] They simplify scaling without duplication of all components. 
\end{enumerate}

To facilitate interoperability of all components and adequate decoupling, one commonly used communication mechanism for microservices is REST (for ``representational state transfer'') \citep{fielding2000}, which uses standard HTTP/HTTPS requests to implement stateless operations to access and manipulate web resources. The resources, designated with an Uniform Resource Identifier (URI), respond to requests with a payload formatted in HTML, JSON, XML, or any other designated format.
Service-oriented approaches have recently been used to implement robotic solutions for assistive tasks. For instance, \cite{napoli2019} presented a robot-as-a-service approach with the intention of customizing the robotic system with respect to specific assistive tasks, and \cite{ercolano2019} introduced a set of socially assistive robot behaviors for monitoring and interaction with elderly people affected by Alzheimer disease, which rely on a microservice architecture currently operating in a private home.
Similarly, OpenTera follows microservices architecture guidelines and extends its usage to broader applications rather than being specific to robotics.
OpenTera's architecture and data representation are designed to address clinical researchers needs by organizing the data collection process in a sites, projects and participants hierarchy. From the caregiver point of view, this architecture allows a patient-centered approach in their daily activities by putting the patient at the center of the software solution by linking directly to it the clinical sessions and related data.
\begin{figure}
\begin{center}
\includegraphics[width=\linewidth]{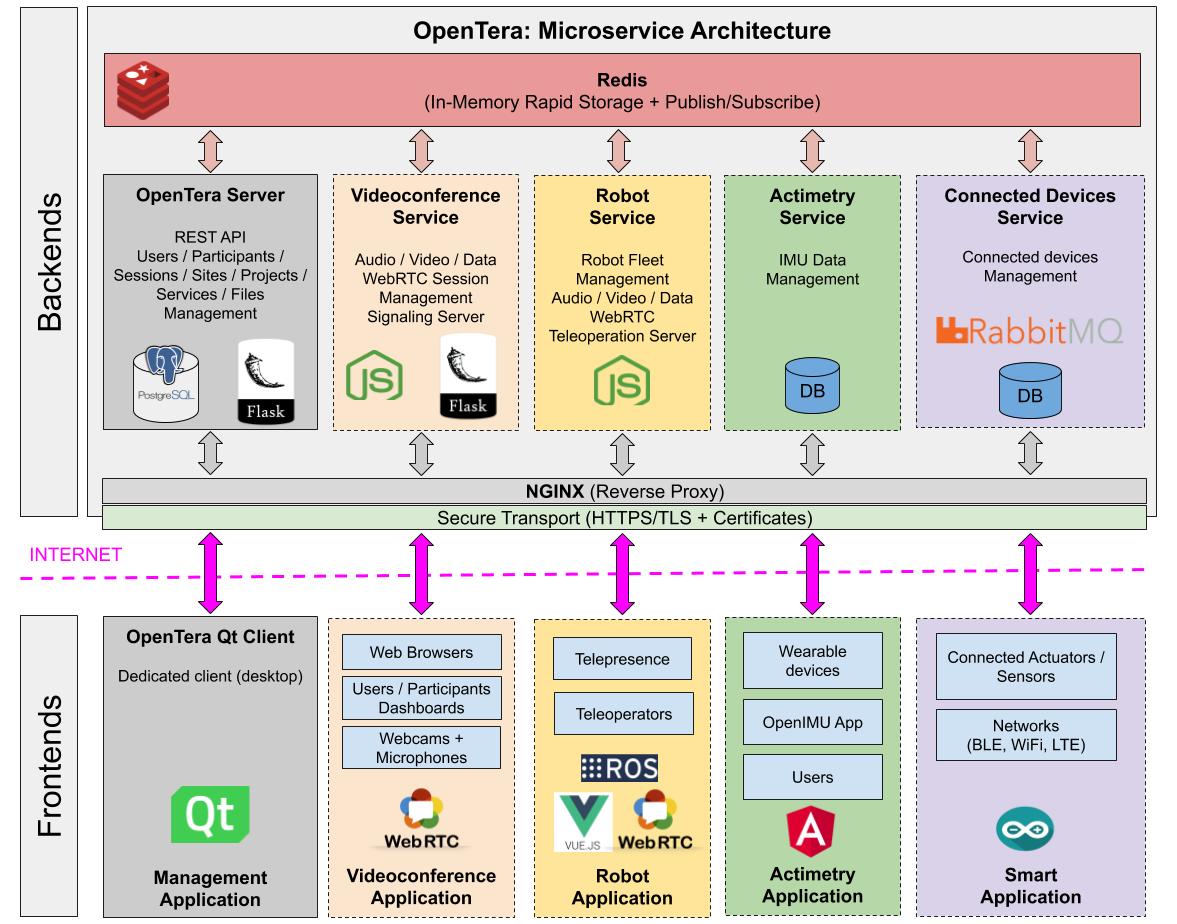}
\end{center}
\caption{OpenTera microservice architecture.}\label{fig:1}
\end{figure}
Figure~\ref{fig:1} illustrates OpenTera's overall architecture, designed to support the development of independent and modular services to facilitate deployment and maintenance.
Using open source packages, the OpenTera architecture integrates its microservices using shared RESTful APIs  and Redis\footnote{\url{https://redis.io}} network messaging technology for internal communication and data storage. 
OpenTera's architecture has two layers: the backends (top) and the frontends (bottom). 
The backends manage the low level of the application (databases, logic, and communications) and work behind the scenes to deliver  information to the user. 
The frontends offer a user-friendly interface, in the form of a web or dedicated application. 
Between the backends and frontends, there is a bridge composed of the NGINX \footnote{\url{https://www.nginx.com}} server and secure transport through HTTPS routes and TLS certificates to protect the information exchanged via the Internet.
The NGINX server acts as a reverse proxy that controls and protects access to services on the private network where the backends reside. Resources appear as if they originated from the proxy server itself, making it visible as a single entry point to the frontend applications. NGINX also handles data encryption and TLS certificates and routes HTTPS requests to the right service.  
Each service offered by the OpenTera architecture is organized to fit its purpose with no dependence on the chosen technology, as long as its interface with other services is satisfied. 
Internally, each service can then implement its own business logic, processing capabilities, and storage to fit the application needs and select the appropriate technologies and programming languages to fit its purpose.
Its main microservices are as follows:
\begin{itemize}
\item \textit{OpenTera Server}: this is a mandatory service that structures users, participants, and device information, and manages access to resources. 
The role of the OpenTera Server is twofold:
\begin{itemize}
\item[--]
Organize and store information in the PostgreSQL database. The server is designed to manage multiple sites, projects, users, user groups, devices, participants, sessions, session types, assets (files, events, session information, observations, forms, etc.), services, and the relations between those data structures to provide data integrity. 
Data creation, retrieval, update, and delete \AQ{\sout{(CRUD)}} is easily accomplished via its REST API using HTTP requests.
This is commonly used in microservice-based systems, since it offers a low-maintenance interface that can easily be implemented in any modern programming language or framework, and it provides decoupling of server components.
In the current implementation, the REST API is implemented with a Python framework called Flask, and the PostgreSQL database is interfaced with SQLAlchemy.
The concept of \textit{session} is used to link all these data structures together.
The session concept stems from a clinical intervention, where it is relevant to keep track of when the consultation occurred, what happened during the intervention, and its duration. Session information can then be consulted again at a later time if necessary.
In OpenTera, a session combines information about which service was used with any user, connected device, and participants, along with timestamped events and data. 
In all cases, the data are only stored in the session if they are pre-authorized by the participants and are subject to prior approval by an ethics committee.

\item[--]
Manage authentication (login) and authorization (permissions) to different parts of the system through its users, participants, devices, and service APIs. 
Authentication can be accomplished by using JSON Web Tokens, SSL/TLS X.509 certificates, or username and password (Http Auth). 
Authorization is normally managed by sites and projects, allowing users, participants, devices, and services to interact with resources that are granted by the site or system administrators. 
\end{itemize}

\item \textit{Videoconference Service}: this provides videoconferencing services for one or multiple audio and video streams. 
WebRTC\footnote{\url{https://webrtc.org}} is the framework used:  it provides real-time communications and uses open standards to send and receive audio, video, and data between peers. %
WebRTC is an open source framework supported by every major web browser through its Javascript interface. 
It allows developers to build web applications that are easy to distribute without any installation of local applications, simply through a web link (facilitating usage by nontechnical users), and works on portable devices (i.e., tablets and smartphones) as well as desktop computers and all operating systems.
The Videoconference Service is designed to offer virtual medical consultations for daily and urgent care and/or doctor visits, virtual communications, and interactions with the family and the community. 
If the virtual call is initiated by the healthcare worker, there is an option of taking notes about the appointment and attaching these to the patient's session. 
The service uses a combination of Python and Flask for its API and Node.js for its WebRTC signaling server, allowing clients to exchange metadata to coordinate communication. 

\item \textit{Robot Service}: 
\AQ{telepresence robots \citep{Kristoffersson2013} are mobile robotic systems, controlled by a human operator via a user interface, which can be used for remote medical check-ups and/or appointments with patients, visits from families, and/or social interaction. 
A recent study \citep{niemela2019} suggests that through telepresence robots, family members are more present for  older adults aging in residential care. 
It is therefore important to include such services in OpenTera, whether these robots are fully teleoperated or having some level of autonomy.
Similarly to the Videoconference Service, the Robot Service} provides videoconferencing and bidirectional data exchange through WebRTC, adding the possibility to bridge control and sensor data from ROS, on the robot side, to the teleoperation interface in the robot application at the frontend level.
In the context of multiple robots in residences, stakeholders need to be able to know the state of all robots in real time, their physical position, their in-use status, their battery state, etc., to make sure that adequate and safe interventions with the robots are performed. 
OpenTera offers  secure access to robots, by managing users' permissions for each robot available in the fleet. Also, it addresses protection of privacy concerns for family members and healthcare workers by connecting the robots without storing or sharing video/audio or any kind of information with a third-party, thus providing data security.
\item \textit{Actimetry Service}: this provides wearable devices to monitor human activity. 
Actimetry is the measurement of people's activity, usually performed using wearable devices such as smart watches or custom hardware with embedded sensors (accelerometers, gyrometers, heartrate sensors, GPS, altimeters, force sensors, etc.). 
Activity processing algorithms (number of steps, active minutes, sleep patterns, etc.), which are often tuned for people with particular physical conditions, gender, and age, give an overview of the minutes spent in several energy expenditure categories, asserted with a combination of sensors, ranging from inactive (sedentary) to very vigorous. 
The OpenTera Server and the Actimetry Service can be used to store and organize data collected periodically and automatically to the service. 
Each device is associated with a site, project, and participant in OpenTera, configured with a unique token or certificate. 
Every upload generates a new session with  associated data that can be retrieved for offline short- or long-time analysis. 
Uploads can be performed via WiFi devices connected to the Internet or through a gateway device.
Automated measurements can also be generated by activity processing algorithms and stored in the session. 
The use of such a nonintrusive service can be very important in detecting changes in usual behavior, which can be used as warnings. 
This information can be displayed in various forms, such as custom dashboards or direct notifications, and can be helpful when planning in-person, videoconference, or teleoperation interventions.
Some of the work for this service has been initiated and should be completed in the near future with the integration of a custom application called OpenIMU.\footnote{\url{https://github.com/introlab/OpenIMU}}
OpenIMU is an open source and generic data importer, viewer, manager, processor, and exporter for inertial measurement units (IMUs) and actimetry data. By using a common sensor data format and structure, data from different sources can be imported and managed in the software using a local database. 
Wearable actimetry devices, once connected to OpenTera and associated with a participant, can automatically and periodically transfer data to sessions. 
We are currently working on an importer that can retrieve raw actimetry data from an OpenTera participant's session.
Once imported, OpenIMU can analyze and produce activity reports that could be added to the participant's session for further consultation.
\item \textit{Connected Devices Service}: this provides real-time bidirectional data exchange with smart home devices. 
Connected devices, often referred as \AQ{IoTs}, comprise a network of heterogeneous computing devices communicating over the Internet to send and receive data without requiring human intervention. Internet communication can be provided by various methods---wired networks or wireless technologies, such as cellular networks (2G to 5G), WiFi, Lora, Bluetooth, or ZigBee---and sometimes require a communication gateway that bridges different protocols and \AQ{provides} interaction of various devices.  
As shown in Figure~\ref{fig:1}, a message broker, such as RabbitMQ \citep{Arasa2019}, which implements standard protocols such as AMPQ (ISO/IEC 19464) and MQTT (ISO/IEC 202922), is used for communication between the devices and with services hosted on the cloud.
International standards are being developed to ensure privacy and security of such systems. Encryption and adequate authentication to the broker is then required, each device having its own and unique identifier and credentials.  
Ongoing development of the service includes two complementary applications with connected devices. 
The first application is related to healthcare applications \citep{ahmadi2019}, with wireless medical devices that are producing data for specific patients or participants. Most of the time, the data acquisition are going to be performed by wireless devices that are sometimes shared in LTC facilities.
For instance, the same thermometers, weight scales, blood pressure, or SpO2 sensors are reused with multiple persons, and data are collected through a single data hub (a robot, computer, tablet, phone, etc.). In this case, it is mandatory that the service dynamically assign the data producing device to the right participant to ensure data confidentiality. This can be configured by residence staff, by automatically detecting the user, or by the elderly, and will require training and adequate user interfaces.
The second application we envision is related to smart environments, which can include smart homes, smart cities, and smart manufacturing, which are augmented with various kind of sensors. 
In the context of LTC for seniors, we define smart homes as residences in which devices (examples of such devices can be found in \cite{rashidi2013}) can be used for a variety of purposes:
\begin{itemize}
\item[--] In shared spaces, they can be used to change TV channels, adjust chair positioning and lighting, and control music preferences.
\item[--] They  can  allow monitoring of  residents. 
Monitoring residents' health can help health workers perform virtual visits.
Collecting environmental data through connected devices is easy, but associating data with a given individual is more challenging in a context where multiple persons use the same environment.
On the other hand, collecting anonymous data can be very useful to optimize use of robots.
\item[--] They can encourage independence and maintenance of good health, and provide pleasurable experiences within residences \citep{lee2020}.
\item[--] They can help in case of emergencies (e.g., through detecting falls by residents or high CO$_2$ levels and by providing emergency contact) and can feed robots with information to display or communicate to residents.
\item[--] They can facilitate robots' navigation throughout the residence, for example, by automatically opening  doors to allow easier movement in the halls.   
\end{itemize}
\end{itemize}
%

\section{Needs and Challenges Identified during the COVID-19 Pandemic}
\label{sect3}

When the COVID-19 lockdown started, our research team, in collaboration with \AQ{clinicians and partners from the health sector}, started to think about useful short-term applications to help address the difficulties observed in healthcare. The following are examples of applications that we identified:
\begin{itemize}
\item Design a virtual patient triage station that would allow a nurse to remotely assess the condition of a potential patient. This would allow older nurses, who can be at risk, to intervene remotely and limit unnecessary movement of infected people. 
\item Use mobile robots for disinfecting or monitoring areas.
\item Control a pan-tilt thermal camera to remotely measure people's temperature. 
\end{itemize}

\AQ{We then reached} out to two LTC facilities in Québec Canada, to gather their ideas about how technology could help them in this time of crisis. 
Hereinafter, the care facilities are referred to as CF1 and CF2. 
First, virtual meetings through online applications (\textit{A-meets}) were conducted, separately with CF1 and CF2, to determine the needs and challenges encountered by each of these facilities. 
Then, additional virtual brainstorm meetings (\textit{B-meets}) followed, again separately with CF1 and CF2, to explore further challenges and needs, as they are ready to conduct trials in early and late Fall 2020. 
The virtual A-meets and B-meets were conducted after the lockdown period of the pandemic, between mid-May and end-August 2020. 
Table~\ref{tab:1} presents a summary of these meetings. 
Five out of ten needs identified were COVID-related.   
Both care facilities raised the following common needs: 
\begin{enumerate}
\item[(a)] constant repetition of the COVID restrictions, namely, \textit{Wash your hands for at least 20 seconds}, \textit{Keep a distance of 2 m}, and \textit{Avoid interaction in groups};
\item[(b)] personalized assistance that can engage patients in social interaction and can help healthcare workers to provide care in accordance with each patient's profile; 
\item[(c)] security tasks, to detect if a person  is lost in the facility.
\end{enumerate}
CF1 also identified two other COVID-related needs:
\begin{enumerate}
\item[(d)] managing who enters and exists the care facility;
\item[(e)] patrolling to check that the older adults do not leave their rooms during sleeping hours.   
\end{enumerate}
Ideas about other needs also emerged during the meetings, such as having access to residents' ambient temperature or providing pill and water reminders, to help give healthcare personnel  time to dedicate themselves to more urgent tasks.  
Having faced the difficulties associated with the pandemic, both care facilities expressed their interest in getting extra help through the use of robotic solutions.  
Sterilizing the robots was their first concern, each facility having different strategies regarding how to proceed: CF1 would only use sprays to avoid spending time using chiffon to clean  up, whereas CF2 would use sterilized wipes or chiffon and sanitizing solutions.
\begin{table}[htbp]
\centering
\captionsetup{justification=centering}
\caption{Potential solutions to needs identified while meeting healthcare workers from care facilities}\label{tab:1}
\begin{tabular}{llllll}
\hline
  & &  &  & & \\
  & Needs & Care facility & Meet & COVID-related & Potential solution \\
  & & (CF1, CF2) &  &  (Yes/No)  & \\
   &&  &  &  &\\ \hline
  N$_1$ & Wellness and health &  & & &  \\ 
   &\multicolumn{1}{r}{\textit{Room temperature}} & CF2 & B & No & IoT \\
  N$_2$ & Tele-assistance &  &  &  & \\
   &\multicolumn{1}{r}{\textit{Doctor's visits}} & CF2 & A & Yes & Telerobots\\
   &\multicolumn{1}{r}{\textit{Assist nurses}} & CF2 & A & Yes & Telerobots \\
  N$_3$ & Constant reminders &  &  &  &\\
   &\multicolumn{1}{r}{\textit{To drink water}} & CF2 & B &  Yes \& No & IoT, SAR \\
   &\multicolumn{1}{r}{\textit{To take medication}} & CF2 & B &  Yes \& No & IoT, SAR\\
  N$_4$ & Logistics &  &  &  & \\
   &\multicolumn{1}{r}{\textit{Move medical supplies}}& CF2 & B  & Yes \& No & Mobile robots\\
  N$_5$ & Social interaction &  &  &  & \\
  & \multicolumn{1}{r}{\textit{Connect with family}} & CF1 & B & Yes & Telerobots, SAR \\
     &\multicolumn{1}{r}{\textit{}} & CF2 & A, B & Yes & Telerobots, SAR\\
   &\multicolumn{1}{r}{\textit{Connect with the community}} & CF2 & A, B & Yes & Telerobots, SAR\\
  N$_6$ &Activities &  &  &  & \\
   &\multicolumn{1}{r}{\textit{Accompanied walks}}& CF2 & B & Yes & SARs, Mobile robots\\
   &\multicolumn{1}{r}{\textit{Engage in social activities}} & CF2 & A, B & Yes & Telerobots, SAR\\
  N$_7$ & Covid-related & &  & & \\
  & \multicolumn{1}{r}{\textit{Questioning}} & CF1, CF2 & A, B & Yes & SAR, loudspeakers, telerobots\\
     &\multicolumn{1}{r}{\textit{Repeat COVID-19 restrictions}} &  CF1 & A, B & Yes & SAR, loudspeakers, telerobots\\
   &\multicolumn{1}{r}{\textit{}} & CF2 & A & Yes & SAR, loudspeakers, telerobots\\
  & \multicolumn{1}{r}{\textit{Take temperature}} & CF1, CF2  & A, B  &Yes & Thermal camera, IoT\\
  N$_8$ & Person and action recognition &  &  &  & \\
   &\multicolumn{1}{r}{\textit{Manage entries/outputs}} & CF1 & A, B & Yes & Any robot with a camera\\
  N$_9$ & Personalized assistance & CF1, CF2 & A, B & Yes \& No & SAR\\ 
  N$_{10}$ & Patrolling & & &  & \\
   &\multicolumn{1}{r}{\textit{Find persons}} & CF1 & A & Yes \& No & Mobile robots\\
   &\multicolumn{1}{r}{\textit{Security}} & CF1, CF2 & A, B & Yes \& No & Mobile robots\\
 \hline
\end{tabular}
\end{table}

A number of possible solutions were discussed during the B-meets, and a number of observations were made:

\begin{itemize}
    \item The use of IoT devices, such as connected thermometers or connected loudspeakers, can be considered as a solution to  N$_1$--Wellness and health and N$_3$--Constant reminders. 
    However, another option could be the use of \textit{companion} robots \citep{broekens2009}, such as MiRo \citep{Prescott2017}, Aibo,\footnote{\url{https://us.aibo.com}} Paro \citep{Wada2007}, iCat \citep{Breemen2005}, and Huggable \citep{Stiehl2006}, %
    which can provide constant reminders or alerts when the environment is not appropriate for  older adults. Such robots can also be considered for  N$_9$--Personalized assistance. Moreover, mobile robot manipulators, such as the Tiago robot, can also be considered for  N$_9$, since a recent study \citep{cosar2020} has already used it to provide personalized assistance to older adults aging in place. 

    \item Telepresence robots, either commercialized ones such as Beam, OhmniLabs, or Giraff \citep{Kristoffersson2013} or modified versions with assistive capabilities \citep{laniel2017, laniel2017b}, can address  N$_5$--Social interaction and allow  older adults to be in touch with their families and the community. 
    Telepresence robots can also be considered for  N$_2$--Tele-assistance  to provide virtual check-ins with doctors and/or nurses. Such robots can be designed to have attached supports, such as plates or baskets, which can be used to carry the tools necessary for taking  vital signs and the usual medical supplies used by  nurses during  check-ins, as required by  N$_4$. 
    
    \item An important identified need is N$_6$--Activities, since such activities are thought to stimulate  older adults'  cognitive functions. Before the imposition of social distancing restrictions, older adults gathered in large groups to interact through social activities. Now, such kinds of physical gatherings are limited. 
    Still, telepresence robots could be used by local organizers to allow them to be present remotely on the robot while coordinating games or other social activities for small groups of older adults.
    Another option is the use of socially assistive robots (SARs), such as the Sprite robot, which has played the role of a moderator in a group interaction centered around a tablet-based assembly game \citep{short2017}.  
    Studies involving teleoperated robots or SARs  stepping up in crisis situations are not new: tests were performed with participants  in a simulated Ebola treatment unit (ETU) while a SAR, the PR2 robot, performed various tasks via teleoperation \citep{kraft2016}. The results suggest that participants trust the robot more when they can see the teleoperator. The solutions presented in this study give  older adults the opportunity to see the operator on the tablet attached on the robot.
    
    \item The N$_7$--COVID-related need can be addressed by using wireless thermometers or thermal cameras \citep{cosar2020} to take people's temperature, and loudspeakers, mobile robots, or SARs to repeat the social distancing guidelines and to ask the standard COVID-19 related questions. 

    \item The N$_8$--Person recognition need to manage who is entering and exiting the care facility can be addressed by using any robot with a camera attached and a proper recognition algorithm. 

    \item N$_{10}$--Patrolling through the care facility, for example at night during sleeping hours, can be performed by autonomous mobile robots with cameras to detect unusual behavior such adults being where they are not supposed to be. Another solution could be a telepresence robot controlled remotely.     

\end{itemize}

\section{How O\lowercase{pen}T\lowercase{era} Can Help}
\label{sect4}
One outcome of the needs and challenges outlined in Section~\ref{sect3} is that rapid prototyping is  essential to address a variety of specific applications and requirements. 
This section outlines how OpenTera can be used to implement identified applications, along with specific functionalities integrated into its microservice architecture.

\subsection{Patient Triage \AQ{Application}}\label{triageapp}
\begin{figure}[htpb]
      \centering
        \includegraphics[width=.45\textwidth]{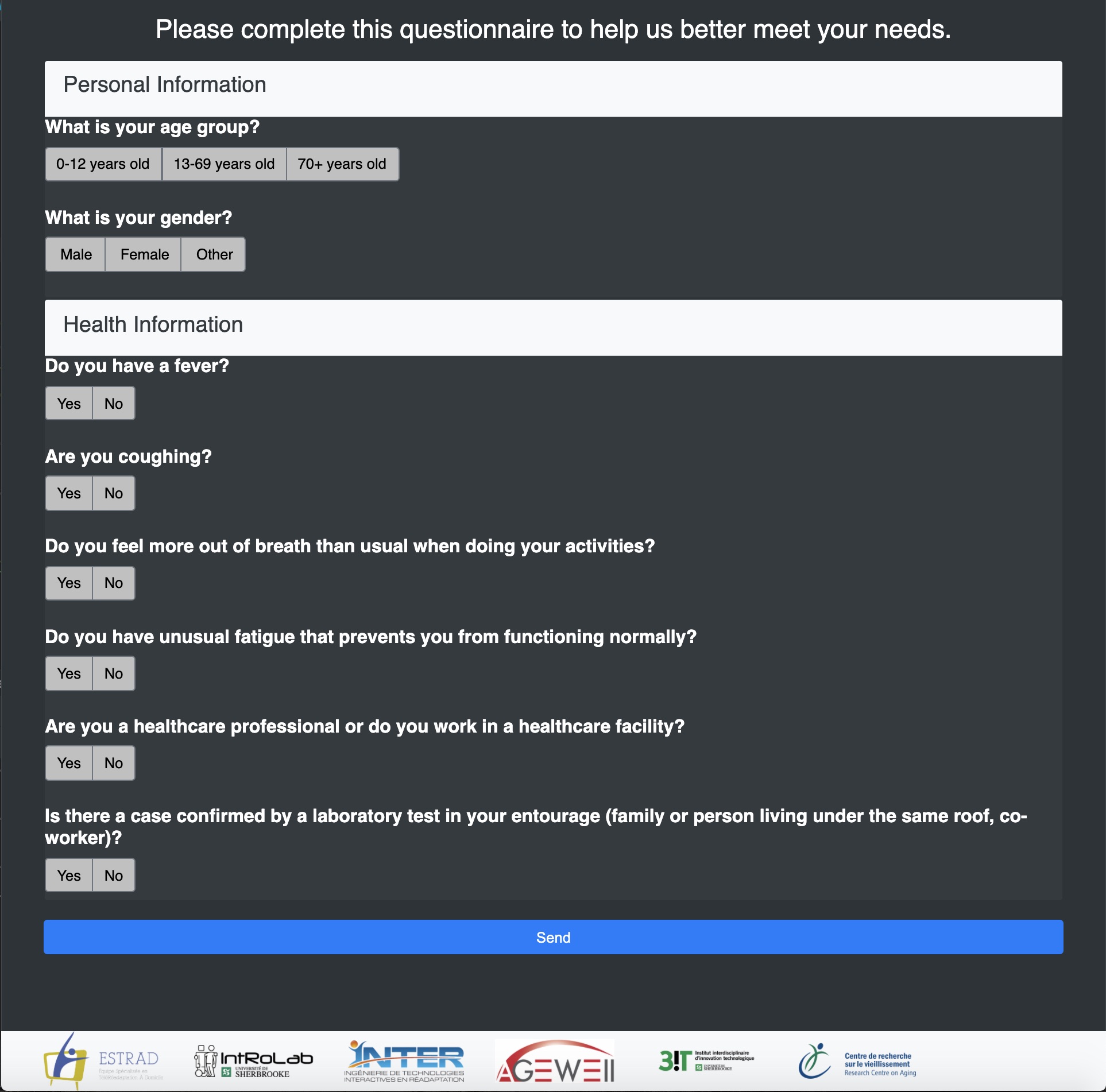}
        \includegraphics[width=.45\textwidth]{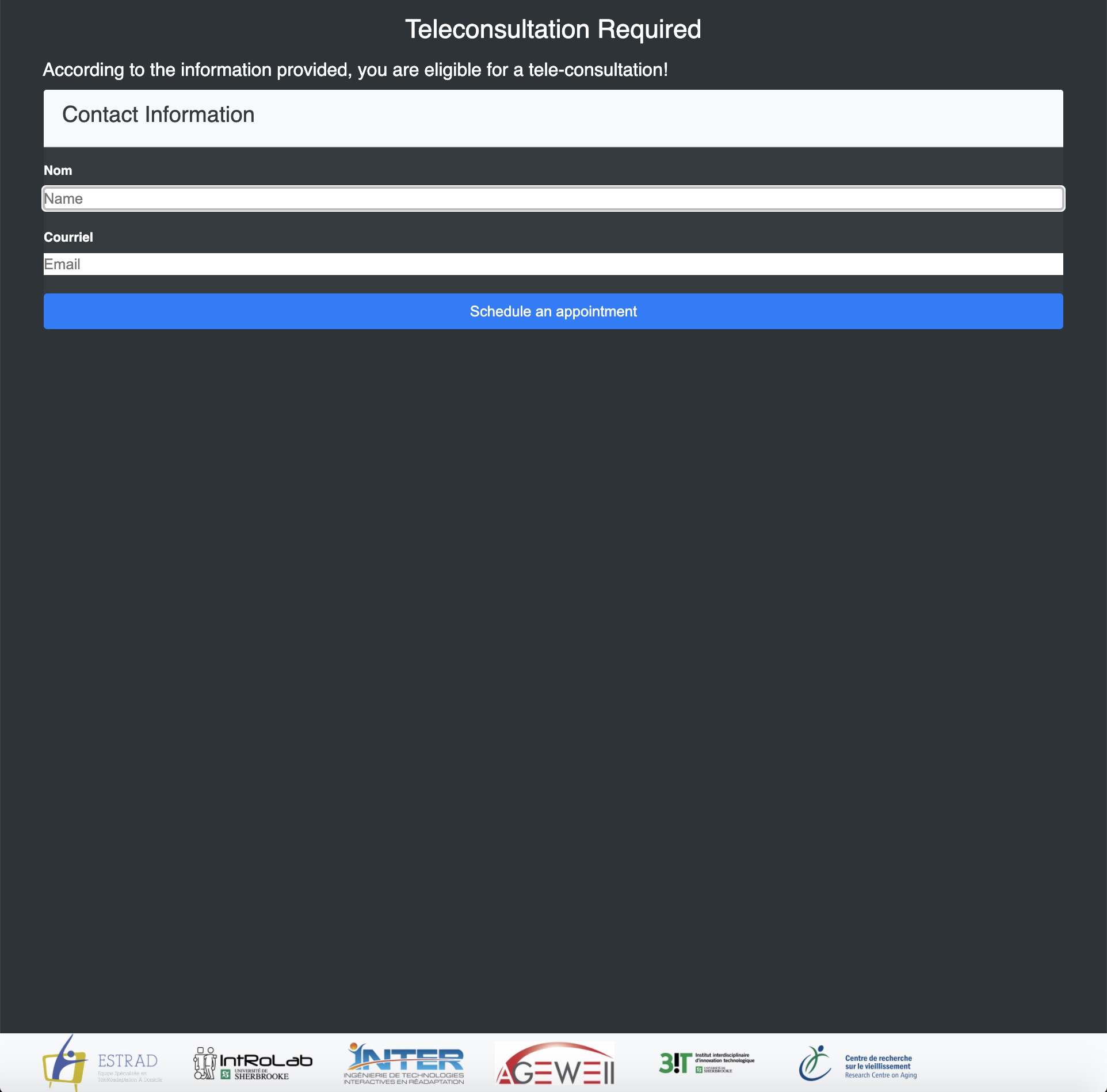}
        \includegraphics[width=.5\textwidth]{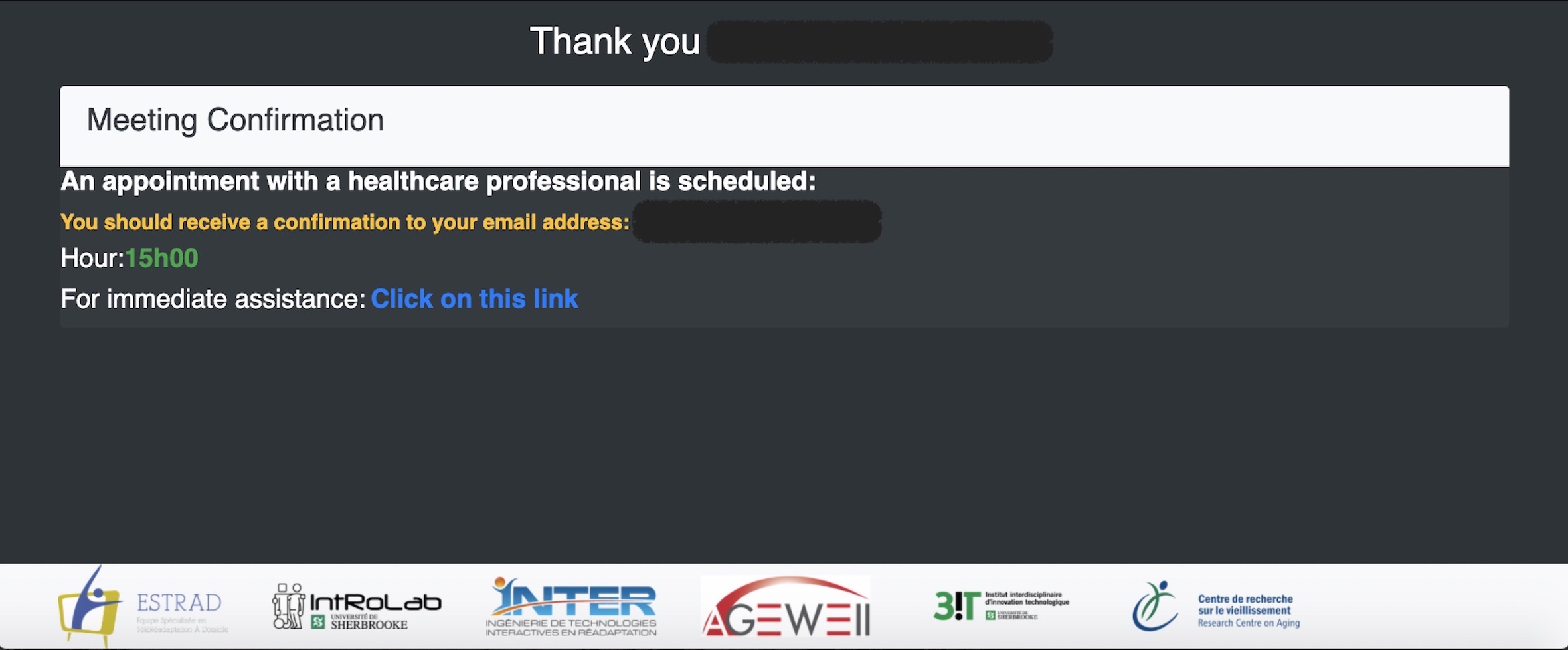}
        \includegraphics[width=.4\textwidth]{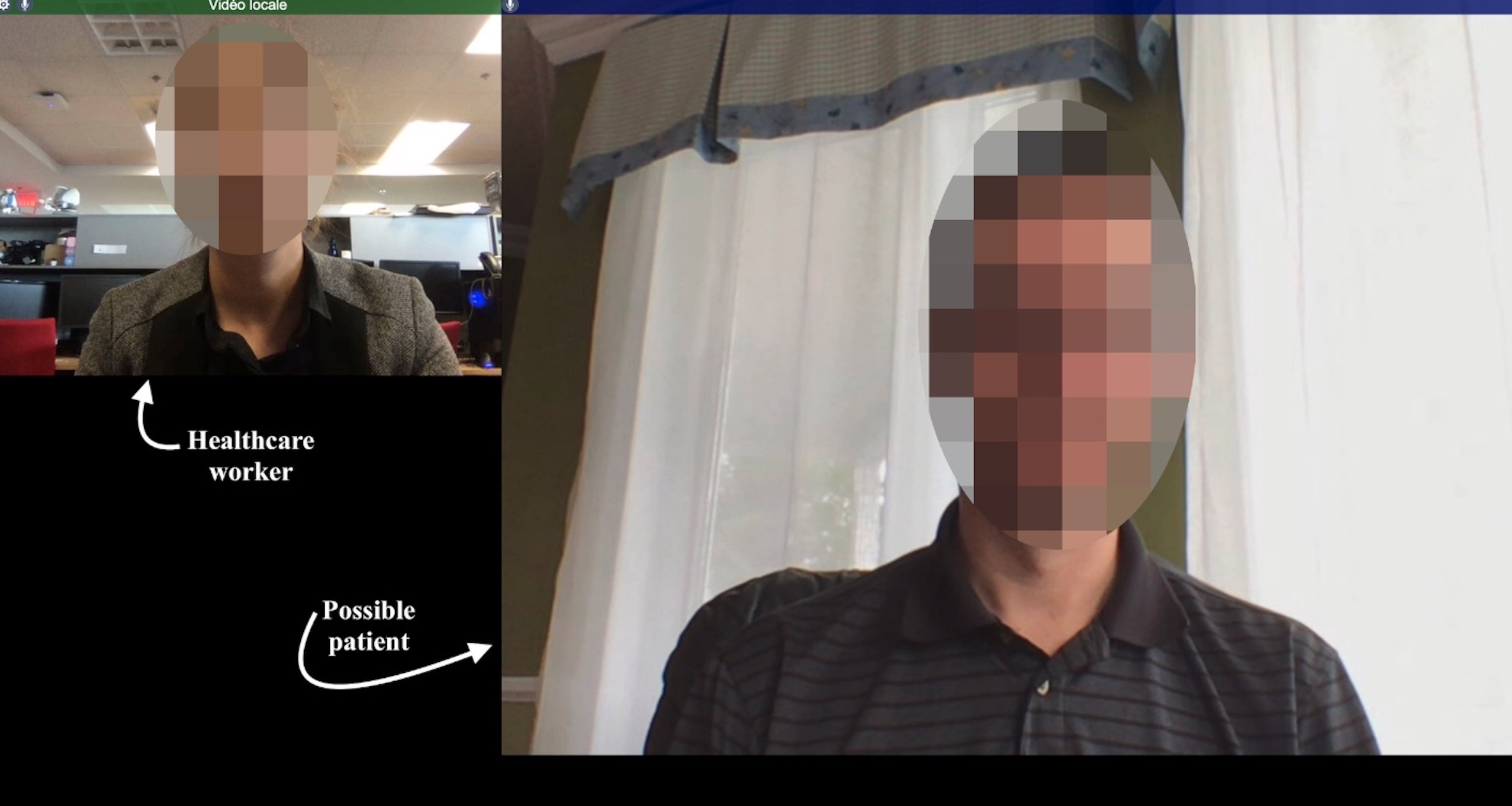}
      \caption{Virtual patient triage application using OpenTera: \textbf{(A)} questionnaire; \textbf{(B)} notification of need for appointment; \textbf{(C)} appointment schedule; \textbf{(D)} Tele-consultation.}\label{fig:triage}
\end{figure}
At the start of the pandemic, a simple patient triage application was prototyped using OpenTera in three weeks by two developers. The application consists of accessing a virtual meeting room using a web link, designed using OpenTera's videoconferencing service module. 
The user is first asked to fill out a questionnaire to identify if important symptoms related to the virus makes it a priority.
The user then waits to have a videoconference session with a nurse, who is going to provide recommendations for what to do next. 
Figure~\ref{fig:triage} illustrates the typical interaction scenario with this application:
\begin{itemize}
    \item Figure~\ref{fig:triage}A: the user starts to fill out personal (\textit{age group}, \textit{gender information}), health (\textit{fever}, \textit{cough}, \textit{tiredness}), and other information (\textit{Is the user a healthcare worker?}, \textit{Has the user been in contact with someone with a confirmed COVID diagnosis?}) on the provided questionnaire. Based on their responses, the user is assigned a teleconsultation with a nurse.  
    \item Figure~\ref{fig:triage}B: to fix an appointment, the user's name and  email address are required so that they can receive the appointment information.
    \item Figure~\ref{fig:triage}C: when the appointment is confirmed, an email is sent to the user indicating the date of the appointment and the link to be used to attend the teleconsultation. 
    \item Figure~\ref{fig:triage}D: at the time set for the appointment, the user clicks on the provided link (also sent  via email) and starts the videoconference with a nurse.  
\end{itemize}

\subsection{In-Home Telerehabilitation Application}\label{telerehabilitationapp}
\begin{figure}[htpb]
      \centering
        \includegraphics[width=.98\textwidth]{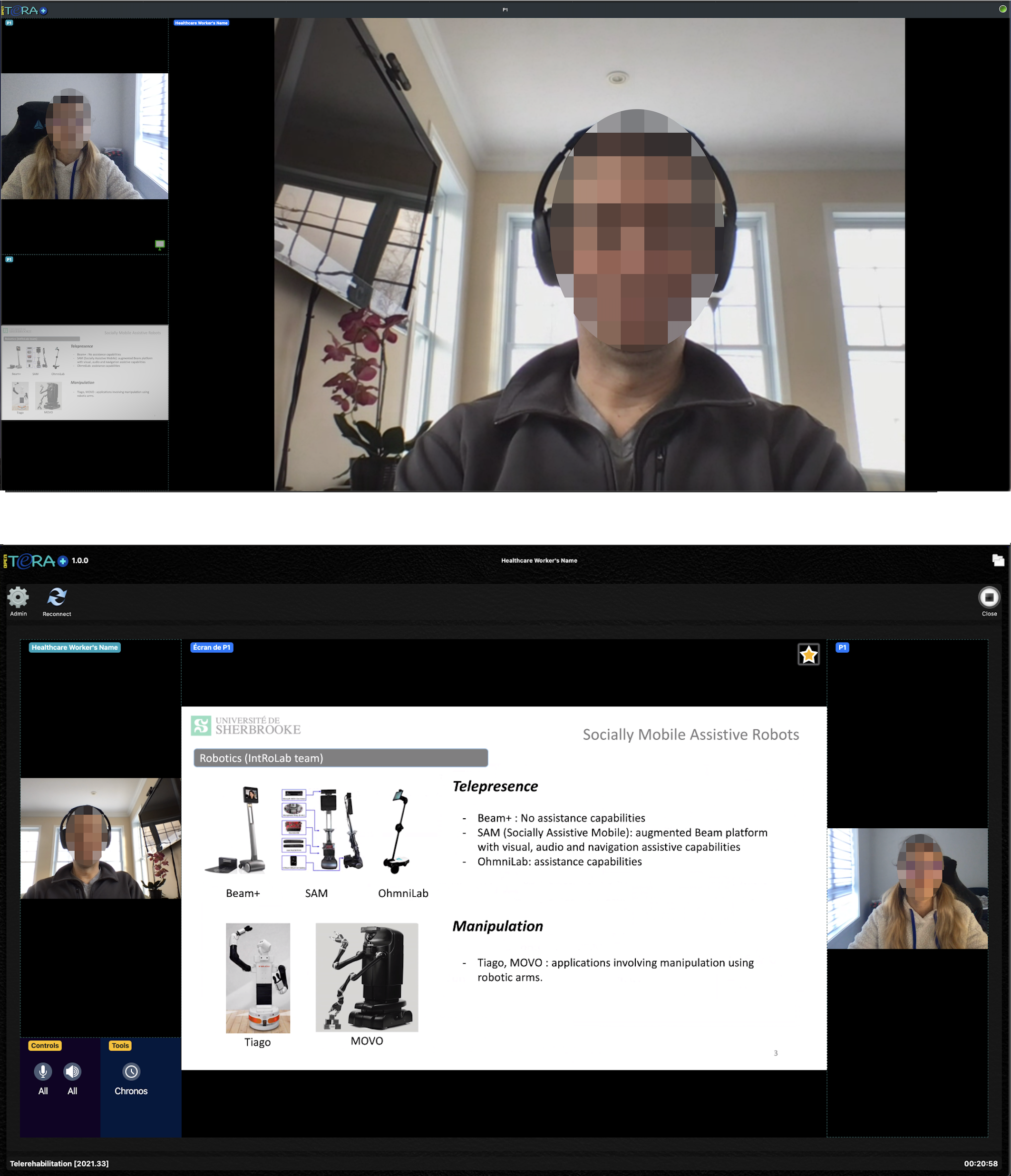}
      \caption{\AQ{Telerehabilitation session interfaces designed with OpenTera: (top) Participant's web-browser; (buttom) Host's desktop application. In this scenario the host shares a PowerPoint presentation. }
      }\label{fig:telerehabilitation_app}
\end{figure}

In-home telerehabilitation is defined as the provision
of remote rehabilitation services in their home \citep{tousignant-moffet-2011}.
A telerehabilitation session involves a healthcare professional connecting to one or multiple patients in their home using audio, video and sensor data. 
A typical session consists of physical rehabilitation including supervised exercises, coaching, training and discussions, but may also involve speech rehabilitation, social and psychological consultations.

Figure \ref{fig:telerehabilitation_app} presents the user-friendly interface for the telerehabiliation session designed using OpenTera, where Figure \ref{fig:telerehabilitation_app}A illustrates the web browser available to the participants, while Figure \ref{fig:telerehabilitation_app}B reports the part of the application only seen by the session host. 
It uses the videoconference service module validated while prototyping the patient triage application, allowing us to reuse previous software components for both the web frontend and the service backend.
A telerehabilitation session is scheduled by having the session host send an invitation to one of multiple participants via email with a web link to connect to the OpenTera telerehabilitation service. 
After having clicked on the link, the participants arrive in a waiting room. 
When ready, the host starts the session and each participant joins it automatically. The functionalities offered by OpenTera's telerehabilitation application are as follows:

\begin{itemize}
    \item Standard videoconference features are provided. The session host has full control of cameras and microphones for all patients and can turn them on/off at any moment, while the participants have full control over their own devices.
    Desktop sharing makes it possible to display any other information by both the host and the participants, such as a PowerPoint presentation for instance (as illustrated in Figure \ref{fig:telerehabilitation_app}).
    The host can add or remove participants from the session at all time using features on the left panel. 
    Giving the opportunity for on-demand group sessions and assistance, a host can invite both patients and healthcare workers as participants to the session.
    The current group limit is five. 
    \item Tools can be used during the sessions, such as a timer to be displayed for each participant during a rehabilitation exercise. 
    We are currently developing a virtual goniometer to measure range of motion of limbs and joints from the video streams. 
    These tools are useful additional features not available on commercial videoconference applications and can greatly influence usability and acceptability of telerehabilitation, as it simplifies and addresses specifically the needs for such application.
    \item The telerehabilitation application makes it possible to manage several sites, users, projects, patients, groups of patients, connected devices, sessions, data sources, and to extract key usage statistics like sessions duration or revisit session events by having access to the session history, for each host and participants. 
%
\end{itemize}

\AQ{
The first version of the telerehabilitation application is available as open source code\footnote{\url{https://github.com/introlab/opentera}}, and is now deployed and used in Paris (France) and Quebec (Canada) for trials and experimentation.
}

\subsection{Telepresence, Mobile Robots, and SARs}

As explained in Section~\ref{sect3}, both CF1 and CF2 indicated telepresence and/or mobile robots as solutions with potentially great benefits during a pandemic.  
\begin{figure}[htpb]
      \centering
        \includegraphics[width=0.28\linewidth]{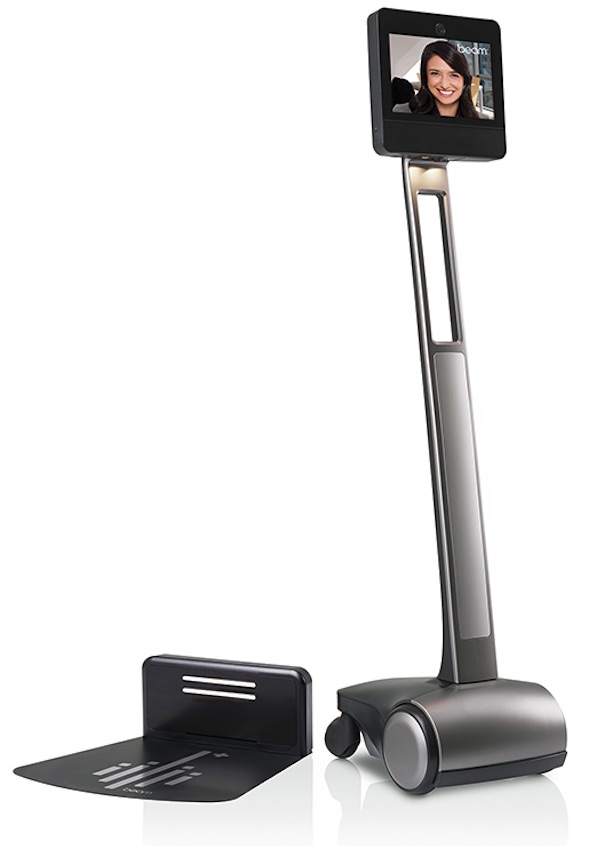}
        \includegraphics[width=0.4\linewidth]{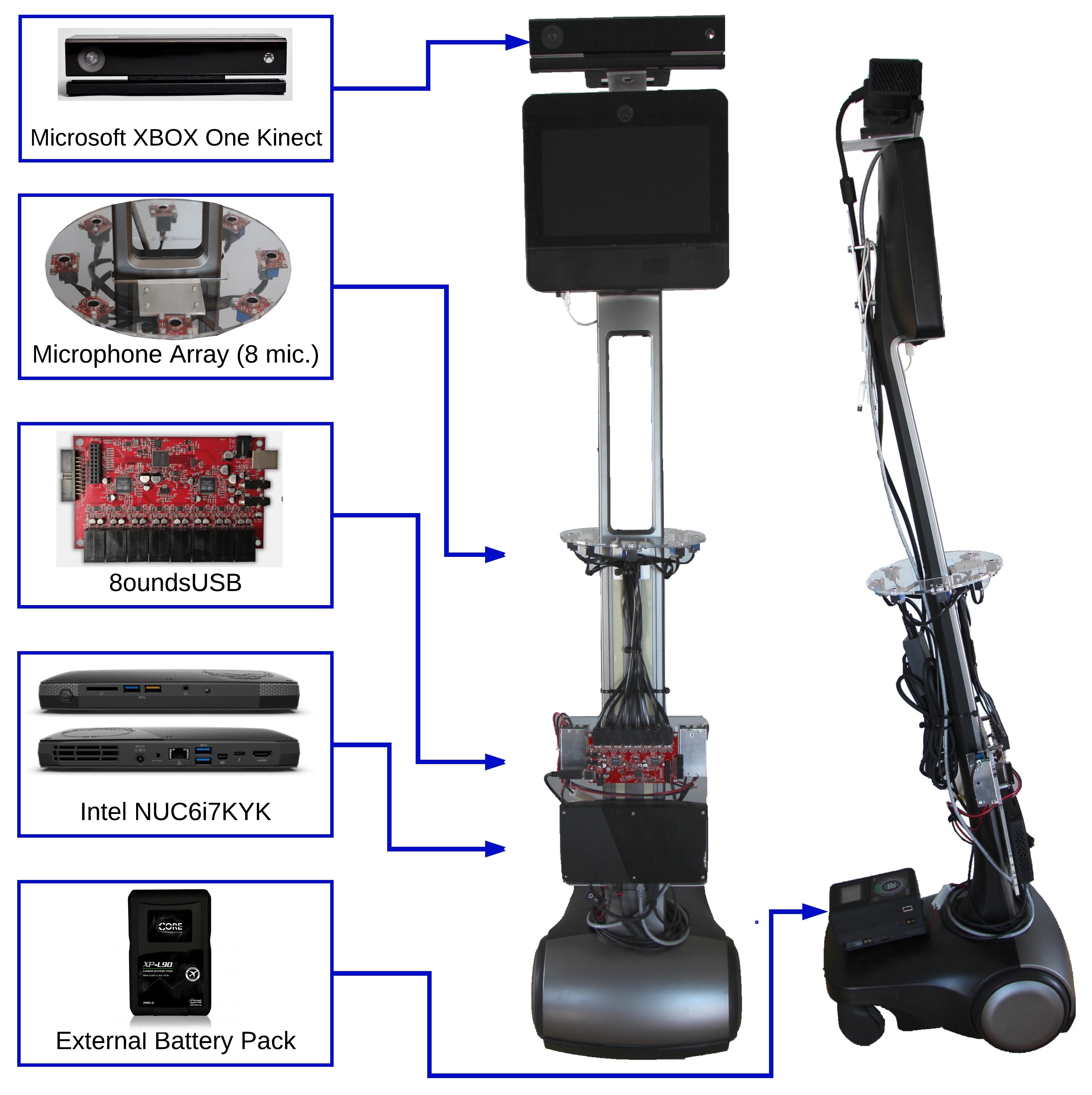}
        \includegraphics[width=0.25\linewidth]{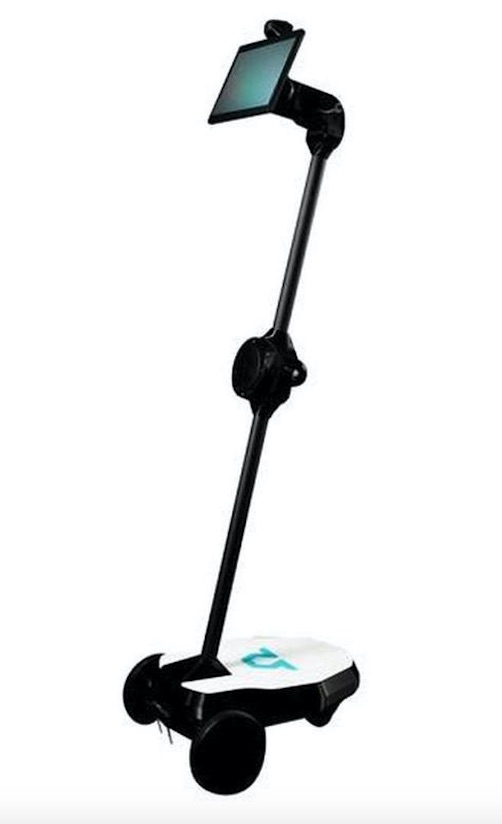}
      \caption{Telepresence robots: \textbf{(A)} Beam; \textbf{(B)} SAM, an augmented version of Beam; \textbf{(C)} OhmniLabs.}\label{fig:2}
\end{figure}
As shown in Figure~\ref{fig:2}, commercial platforms such as Beam and OhmniLabs are available in our laboratory for trials.  
\AQ{OpenTera can be quite useful} in these trials, allowing users to control both platforms using the same telepresence interface and allowing developers to add new features and study their impacts. 
OpenTera can be used on commercial platforms as long as we can have access to a software development kit (SDK) \citep{Benso2000}. Otherwise the platforms must be reverse-engineered and modified to allow custom applications.
We have already augmented a Beam platform that we have named SAM \citep{laniel2017, laniel2017b}, with an RGB-D camera and an additional onboard computer to use Real-Time Appearance-Based Mapping (RTAB-Map)\footnote{\url{http://wiki.ros.org/rtabmap_ros}} \citep{labbe2019}, an open source library for \AQ{simultaneous localization and mapping (SLAM)}.  
We have also added a microphone array for sound source localization, tracking and separation using ODAS  \citep{grondin2019}, which is also an open source library.
Such capabilities are designed to facilitate the teleoperation of the platform by allowing it to navigate autonomously to rooms or back to its charging station (when a session ends, voluntarily or not), \AQ{to orient itself in the direction of the loudest sound source to face the person who speaks, and to avoid colliding with people or obstacles. Ultimately, we want the teleoperator to concentrate on their task (caregiving/rehabilitation/family discussion/medical assessment, etc.) and not on driving the robot around.}
These open source libraries can be used through the frontends applications provided by OpenTera.

\begin{figure}[htpb]
\begin{center}
\includegraphics[width=\linewidth]{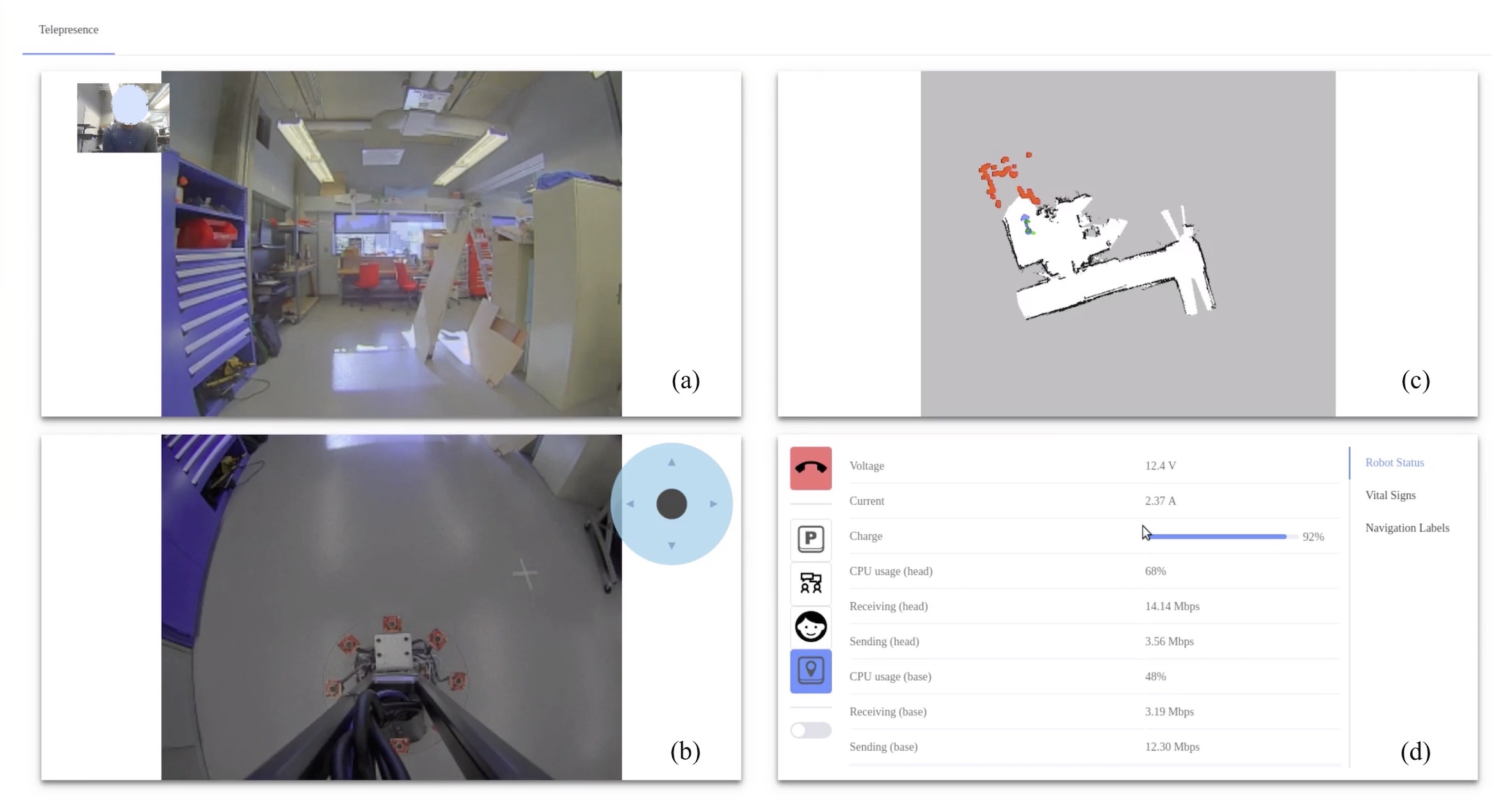}
\end{center}
\caption{Prototype interface of a robot teleoperation service web interface, used by the SAM robot.}\label{fig:robot_teleop_service}
\end{figure}
Figure~\ref{fig:robot_teleop_service} shows a prototype interface available on  OpenTera's frontend layer for the SAM robot, which contains the following: 
\begin{itemize}
    \item Figure~\ref{fig:robot_teleop_service}A and \ref{fig:robot_teleop_service}B:  video images provided by the top and bottom cameras installed on the robot;
    \item Figure~\ref{fig:robot_teleop_service}B: a virtual joystick for robot teleoperation as an overlay to the bottom-facing camera image;
    \item Figure~\ref{fig:robot_teleop_service}C: an interactive map generated by the RTAB-Map algorithm allowing visualization of the laser range sensors (red dots) and waypoint specification (blue arrow) by clicking on the map;
    \item Figure~\ref{fig:robot_teleop_service}D: a window that can provide information about the robot's settings (battery level, CPU usage), vital signs information provided by the IoTs attached to the SAM, and navigation labels (which allow selection of a target destination).
\end{itemize}

\AQ{We are currently using OpenTera as a common base to compare a standard Beam platform, SAM and an OhmniLabs robot, observing usability and acceptability issues that arise during installation and use of the telepresence robot at a residence. Targeted applications consist of using the robot to:
    \begin{itemize}
    \item assist a doctor in carrying out checkups together with a nurse;
    \item allow family members to visit residents and  allow residents to engage in social activities;
    \item carry out patrols;
    \item help nurses to remind residents to take pills or drink water and to remind them about  the COVID-19 instructions;
    \item help nurses to quickly check on residences during the day.
    \end{itemize} 
 }

\begin{figure}[htpb]
      \centering
        \includegraphics[width=\linewidth]{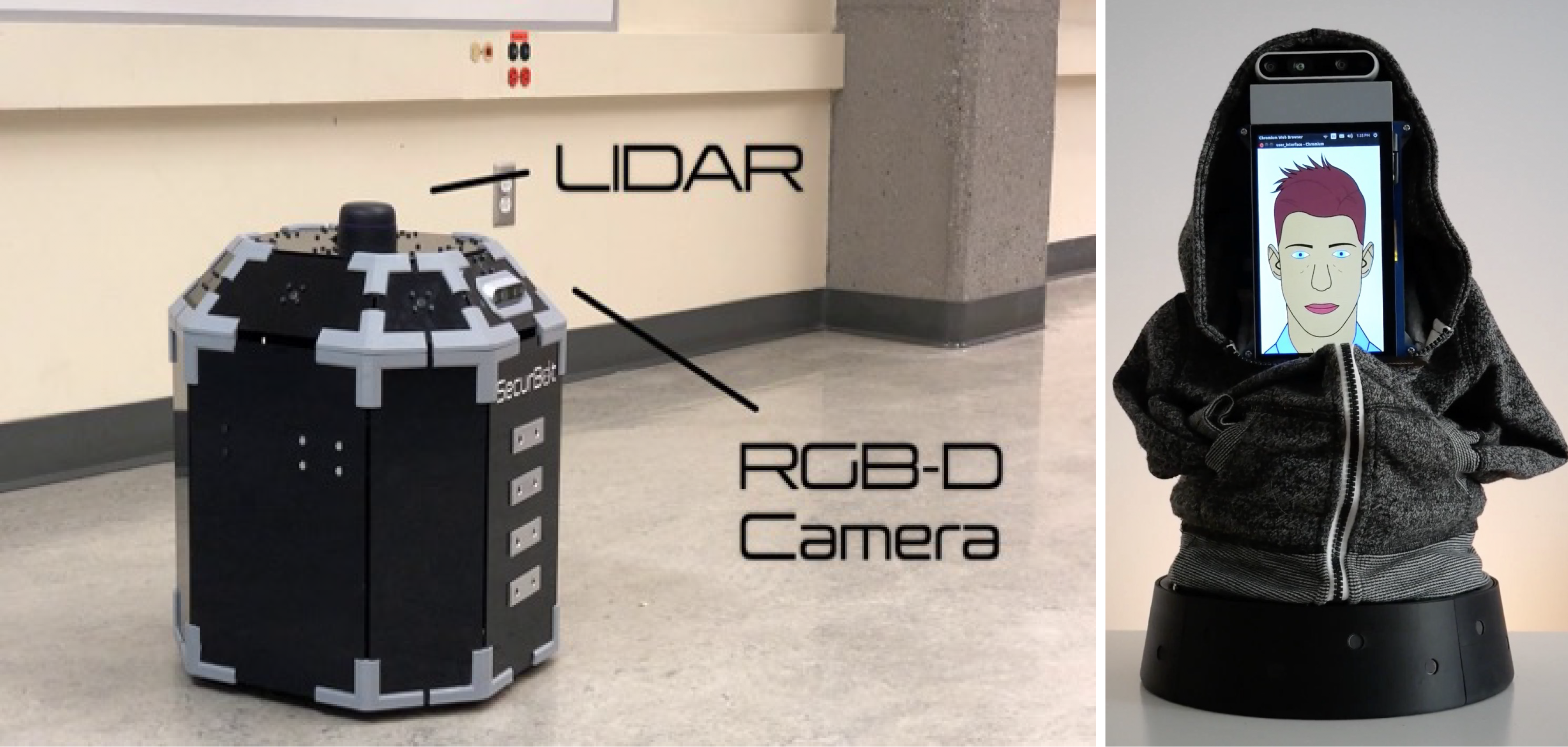}
      \caption{Custom-designed mobile robots and SAR interfaced with OpenTera:  \textbf{(A)} SecurBot; \textbf{(B)} T-Top.}\label{fig:custom}
\end{figure}
Figure~\ref{fig:custom} shows two other custom-designed robots interfaced with OpenTera: 
\begin{itemize}
    \item SecurBot\footnote{\url{https://github.com/introlab/securbot}} is an open source platform with autonomous navigation, person recognition (in case an older adult gets lost in the facility), and other assistive capabilities\AQ{. SecurBot will be used for communicating COVID-19 instructions in a friendly manner and for patrolling}; 
    \item T-Top, a small SAR consisting of a touchscreen mounted on a Stewart platform \citep{Siciliano2008}. Such a robot could act as a COVID-19 assistant, with the duty of managing the residence's entrances and exits. \AQ{T-Top will be enhanced to manage the residence's entrances by having people answer specific questions about COVID-19 (based on the triage application presented in Section~\ref{triageapp}) and by confirming that the person used the hand sanitizer before entering the building. This could be done by interfacing an electronic dispenser and  monitoring the sound it makes when used. If the hand sanitizer is not used, T-Top will immediately contact the reception desk, activating the videoconferencing capability through OpenTera.}
\end{itemize}

Both have access to RTAB-Map and ODAS through OpenTera, facilitating reuse of such capabilities over multiple platforms. 
Such sharing is quite beneficial when telepresence robotic solutions are being deployed in the real world. 
For instance, one of the most common problems encountered when using wireless telecommunication networks is loss of the WiFi signal while navigating a robot. Sometimes, communication can be restored, but in some cases the robot remains jammed in a WiFi dead zone. This unfortunate situation can be remedied by having the robot navigate autonomously, using RTAB-Map for instance, to a place with better signal and wait for the operator to reconnect. But still, evaluating this before deployment in real environments becomes an important step, since it can influence performance and incidentally user adoption.
As a solution, by using OpenTera as an interface to RTAB-Map, we are currently adding WiFi signal strength data to the locations memorized in RTAB-Map,\footnote{\url{http://wiki.ros.org/rtabmap_ros/Tutorials/WifiSignalStrengthMappingUserDataUsage}} thus allowing the user to identify Internet dead zones or areas with weak WiFi signals and have the robot navigate autonomously to locations with better WiFi reception. Moreover, such information can be used to evaluate Internet coverage in care facilities or business offices.
\begin{figure}[htpb]
      \centering
        \includegraphics[width=1\linewidth]{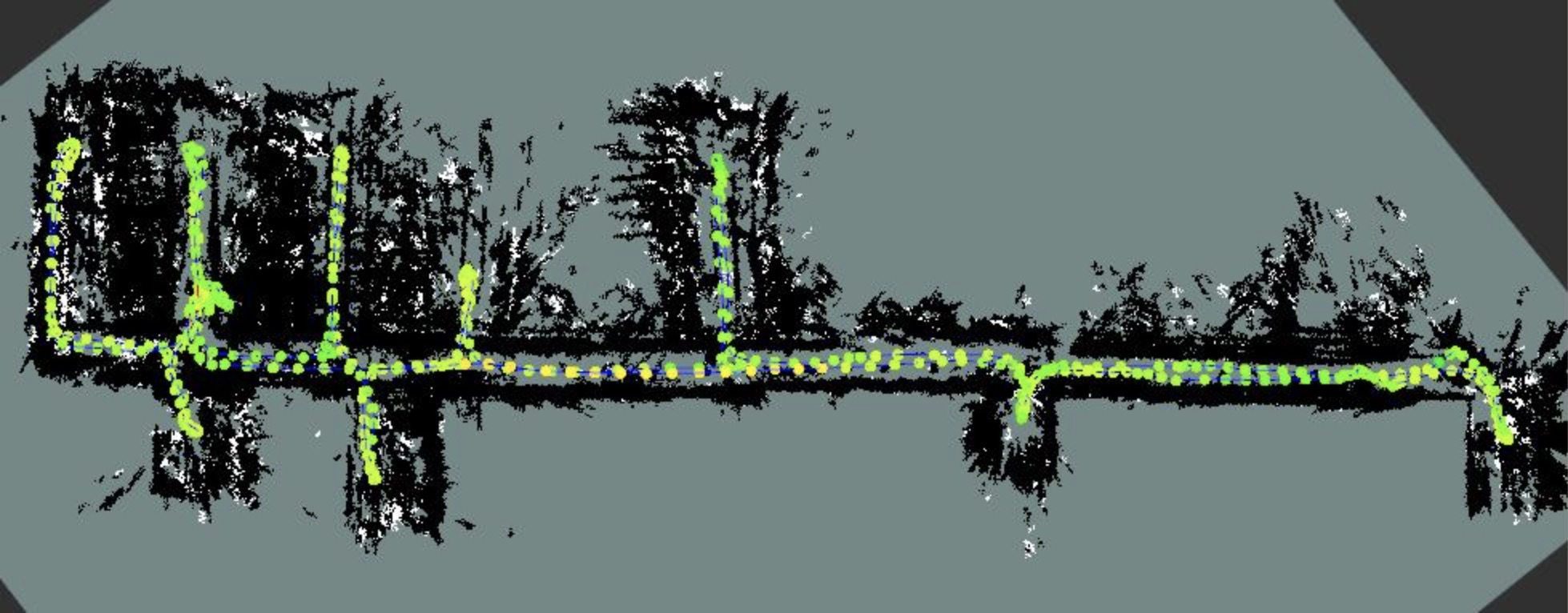}
        \includegraphics[width=0.56\linewidth]{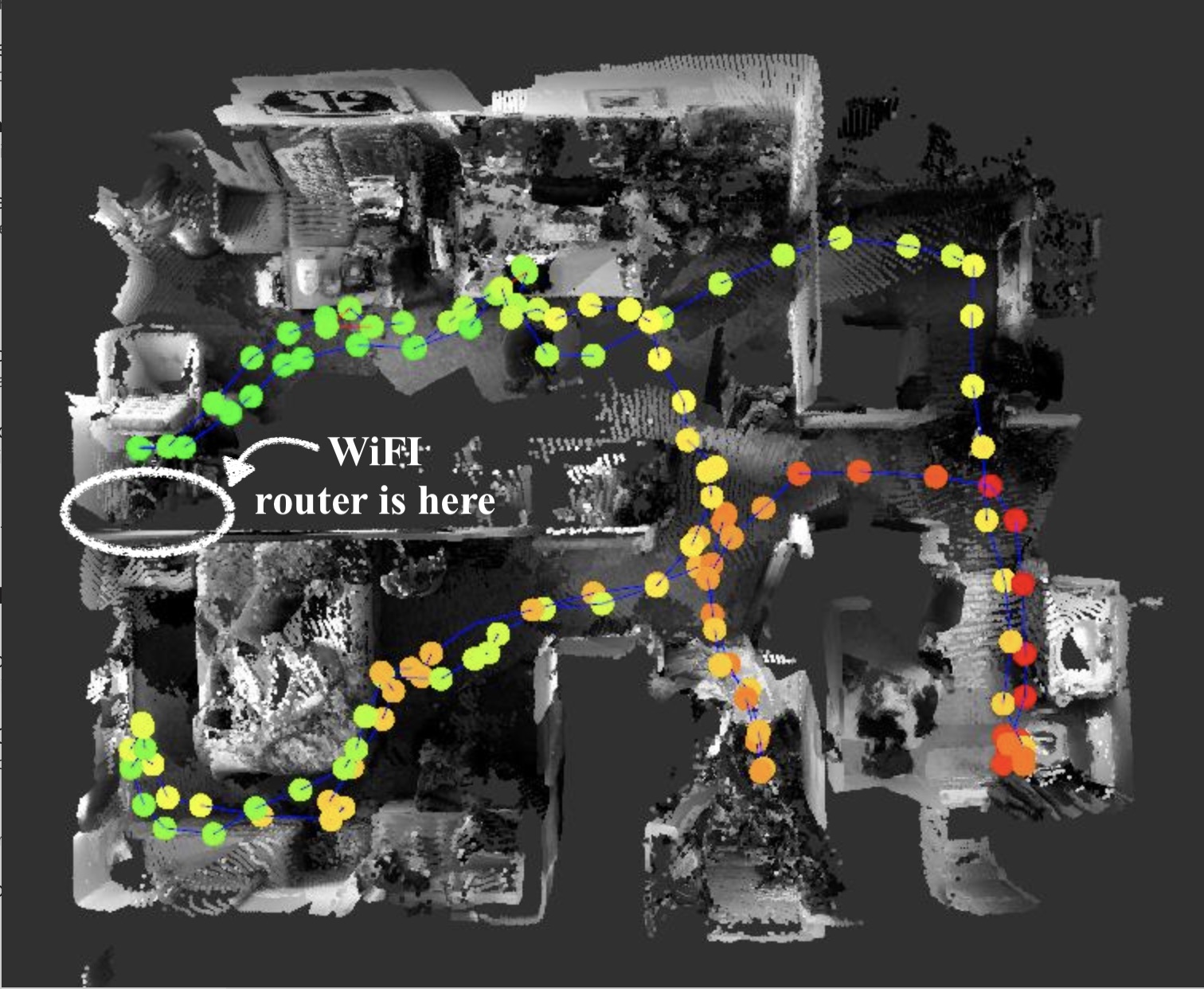}
        \includegraphics[width=0.42\linewidth]{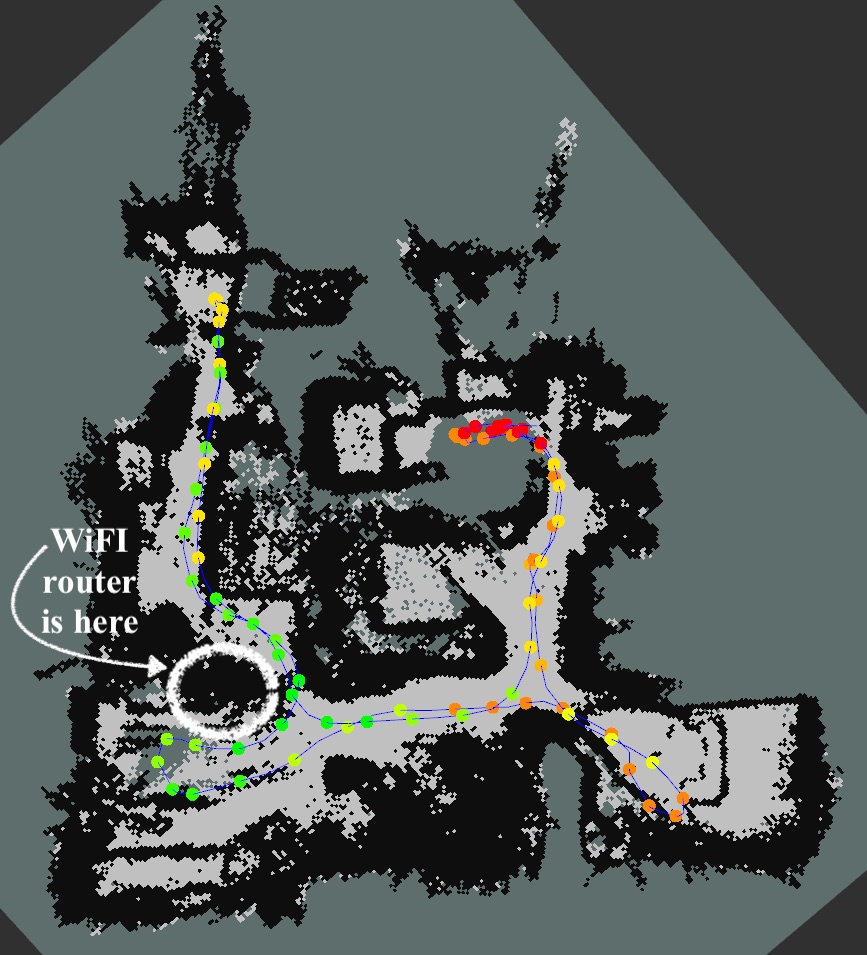}
      \caption{Example of three different environments mapped using the RTAB-Map with WiFi signal strength data: \textbf{(A)} laboratory floor mapped using an Intel Realsense T265 camera; \textbf{(B)} private apartment mapped using a Google Tango ASUS Zenfone AR; \textbf{(C)} private apartment mapped using an Intel Realsense D435 camera.}\label{fig:3}
\end{figure}
Figure~\ref{fig:3} shows how RTAB-Map with WiFi signal strength data works when mapping three environments using different devices. 
The laboratory floor has very good Internet coverage, as indicated by the green dots. 
In the private apartments, areas near the routers have good Internet coverage, but this deteriorates farther away from the routers, as indicated by the orange to red dots. 
Having access to such capabilities through OpenTera would facilitate adaption to specific conditions that may occur. For instance, in a study using the commercially available Double robot \citep{niemela2019}, confidentiality issues were raised by healthcare workers: calls from families could  be performed only in the older adults' room to avoid having family members hear private conversations between other residents or between nurses. Using RTAB-Map through OpenTera for implementing such a capability would be beneficial\AQ{, allowing a community of developers using OpenTera to benefit from this capability in their own designs without having to reimplement everything on their own.}

\section{Conclusion}\label{sect5}
In the context of LTC for seniors in care facilities, this paper explains why a rapid prototyping framework such as OpenTera is important in addressing COVID-19 challenges, and outlines how OpenTera microservices can be used to do a variety of tasks, such as
\begin{itemize}
\item remote evaluation, diagnosis, monitoring, and treatment;
\item provision of tele-assistance services to protect healthcare workers and older adults during crisis situations; \item provision of wearable/H-IoT technologies for monitoring health or specific symptoms;  
\item design of assistive devices to allow older adults to communicate with relatives and healthcare workers using videoconferencing; 
\item mental health support through provision of  providing social robotics and cognitive support through reminders and notifications. 
\end{itemize}

Our objective is to make OpenTera an adaptable solution intended to be used in different robotics/H-IoT applications. 
The examples provided in this paper outline OpenTera's design process and the envisioned testbeds that are to be made available, with the aim of getting others in the \AQ{software design and} robotics community involved in the effort of developing OpenTera, so that \AQ{we can all focus and mutually benefit from} designing innovative solutions to address COVID-19 challenges \AQ{healthcare}.
 
\section*{Conflict of Interest Statement}
The authors declare that the research was conducted in the absence of any commercial or financial relationships that could be construed as a potential conflict of interest.

\section*{Author Contributions}
All authors took part in OpenTera's design and evaluation process. All authors contributed to manuscript revision, and all read and approved the submitted version.

\section*{Funding}
This work is supported by AGE-WELL, the Network of Centres of Excellence of Canada on Aging Gracefully across Environments using Technology to Support Wellness, Engagement, and Long Life, and INTER, FRQNT Québec Strategic Network on Engineering Interactive Technologies for Rehabilitation.

\section*{Acknowledgments}
The authors would like to thank the healthcare personnel and administration staff of the two LTC facilities and homes  for the time they put aside to have meetings with us during this busy and challenging period.     

\bibliographystyle{frontiersinSCNS_ENG_HUMS} 
\bibliography{for_arxiv}
%
%
\end{document}